\documentclass[longauth,letter]{aa} 

\usepackage{graphicx}
\usepackage{amsmath}
\usepackage[varg]{txfonts}
\usepackage{longtable}
\usepackage{url}
\usepackage[citecolor=blue,colorlinks=true]{hyperref}

\newcommand{\Mdepth}   {$20.4 ^{+3.2}_{-3.3}$} 
\newcommand{\Adepth}   {$21.4 ^{+3.3}_{-3.4}$} 
\newcommand{\Ag}   {$0.096 \pm 0.016$} 

\begin{document}

\title{CHEOPS geometric albedo of the hot Jupiter HD\,209458\,b\thanks{The raw and detrended photometric time-series data are available in electronic form at the CDS via anonymous ftp to cdsarc.u-strasbg.fr (130.79.128.5) or via \url{http://cdsweb.u-strasbg.fr/cgi-bin/qcat?J/A+A/}}}

\titlerunning{Geometric albedo of HD\,209458\,b}


\author{A.~Brandeker\inst{1}\thanks{alexis@astro.su.se} \and
K.~Heng\inst{2,3} \and
M.~Lendl\inst{4} \and
J.~A.~Patel\inst{1} \and
B.~M.~Morris\inst{2} \and
C.~Broeg\inst{2,5} \and
P.~Guterman\inst{6,7} \and
M.~Beck\inst{4} \and
P.~F.~L.~Maxted\inst{8} \and
O.~Demangeon\inst{9,10} \and
L.~Delrez\inst{11,12} \and
B.-O.~Demory\inst{2} \and
D.~Kitzmann\inst{2} \and
N.~C.~Santos\inst{9,10} \and
V.~Singh\inst{13} \and
Y.~Alibert\inst{5} \and
R.~Alonso\inst{14,15} \and
G.~Anglada\inst{16,17} \and
T.~Bárczy\inst{18} \and
D.~Barrado y Navascues\inst{19} \and
S.~C.~C.~Barros\inst{9,10} \and
W.~Baumjohann\inst{20} \and
T.~Beck\inst{5} \and
W.~Benz\inst{2,5} \and
N.~Billot\inst{4} \and
X.~Bonfils\inst{21} \and
G.~Bruno\inst{5} \and
J.~Cabrera\inst{22} \and
S.~Charnoz\inst{23} \and
A.~Collier Cameron\inst{24} \and
C.~Corral van Damme\inst{25} \and
Sz.~Csizmadia\inst{22} \and
M.~B.~Davies\inst{26} \and
M.~Deleuil\inst{6} \and
A.~Deline\inst{4} \and
D.~Ehrenreich\inst{4} \and
A.~Erikson\inst{22} \and
J.~Farinato\inst{13} \and
A.~Fortier\inst{2,5} \and
L.~Fossati\inst{20} \and
M.~Fridlund\inst{27,28} \and
D.~Gandolfi\inst{29} \and
M.~Gillon\inst{11} \and
M.~Güdel\inst{30} \and
S.~Hoyer\inst{6} \and
K.~G.~Isaak\inst{31} \and
L.~Kiss\inst{32,33} \and
J.~Laskar\inst{34} \and
A.~Lecavelier des Etangs\inst{35} \and
C.~Lovis\inst{4} \and
A.~Luntzer\inst{36} \and
D.~Magrin\inst{13} \and
V.~Nascimbeni\inst{13} \and
G.~Olofsson\inst{1} \and
R.~Ottensamer\inst{36} \and
I.~Pagano\inst{37} \and
E.~Pallé\inst{14} \and
G.~Peter\inst{38} \and
G.~Piotto\inst{13,39} \and
D.~Pollacco\inst{3} \and
D.~Queloz\inst{40,41} \and
R.~Ragazzoni\inst{13,39} \and
N.~Rando\inst{25} \and
H.~Rauer\inst{22,42,43} \and
I.~Ribas\inst{16,17} \and
G.~Scandariato\inst{37} \and
D.~Ségransan\inst{4} \and
A.~E.~Simon\inst{5} \and
A.~M.~S.~Smith\inst{22} \and
S.~G.~Sousa\inst{9} \and
M.~Steller\inst{20} \and
Gy.~M.~Szabó\inst{44,45,46} \and
N.~Thomas\inst{5} \and
S.~Udry\inst{4} \and
V.~Van Grootel\inst{12} \and
N.~Walton\inst{47} \and
D.~Wolter\inst{22}
}

\authorrunning{Brandeker et al.}

\institute{\label{inst:1}Department of Astronomy, Stockholm University, AlbaNova University Center, 10691 Stockholm, Sweden \and
\label{inst:2}Center for Space and Habitability, Gesellsschaftstrasse 6, 3012 Bern, Switzerland \and
\label{inst:3}Department of Physics, University of Warwick, Gibbet Hill Road, Coventry CV4 7AL, United Kingdom \and
\label{inst:4}Observatoire Astronomique de l'Université de Genève, Chemin Pegasi 51, Versoix, Switzerland \and
\label{inst:5}Physikalisches Institut, University of Bern, Gesellsschaftstrasse 6, 3012 Bern, Switzerland \and
\label{inst:6}Aix Marseille Univ, CNRS, CNES, LAM, 38 rue Frédéric Joliot-Curie, 13388 Marseille, France \and
\label{inst:7}Division Technique INSU, BP 330, 83507 La Seyne cedex, France \and
\label{inst:8}Astrophysics Group, Keele University, Staffordshire, ST5 5BG, United Kingdom \and
\label{inst:9}Instituto de Astrofisica e Ciencias do Espaco, Universidade do Porto, CAUP, Rua das Estrelas, 4150-762 Porto, Portugal \and
\label{inst:10}Departamento de Fisica e Astronomia, Faculdade de Ciencias, Universidade do Porto, Rua do Campo Alegre, 4169-007 Porto, Portugal \and
\label{inst:11}Astrobiology Research Unit, Université de Liège, Allée du 6 Août 19C, B-4000 Liège, Belgium \and
\label{inst:12}Space sciences, Technologies and Astrophysics Research (STAR) Institute, Université de Liège, Allée du 6 Août 19C, 4000 Liège, Belgium \and
\label{inst:13}INAF, Osservatorio Astronomico di Padova, Vicolo dell'Osservatorio 5, 35122 Padova, Italy \and
\label{inst:14}Instituto de Astrofisica de Canarias, 38200 La Laguna, Tenerife, Spain \and
\label{inst:15}Departamento de Astrofisica, Universidad de La Laguna, 38206 La Laguna, Tenerife, Spain \and
\label{inst:16}Institut de Ciencies de l'Espai (ICE, CSIC), Campus UAB, Can Magrans s/n, 08193 Bellaterra, Spain \and
\label{inst:17}Institut d'Estudis Espacials de Catalunya (IEEC), 08034 Barcelona, Spain \and
\label{inst:18}Admatis, 5. Kandó Kálmán Street, 3534 Miskolc, Hungary \and
\label{inst:19}Depto. de Astrofisica, Centro de Astrobiologia (CSIC-INTA), ESAC campus, 28692 Villanueva de la Cañada (Madrid), Spain \and
\label{inst:20}Space Research Institute, Austrian Academy of Sciences, Schmiedlstrasse 6, A-8042 Graz, Austria \and
\label{inst:21}Université Grenoble Alpes, CNRS, IPAG, 38000 Grenoble, France \and
\label{inst:22}Institute of Planetary Research, German Aerospace Center (DLR), Rutherfordstrasse 2, 12489 Berlin, Germany \and
\label{inst:23}Université de Paris, Institut de physique du globe de Paris, CNRS, F-75005 Paris, France \and
\label{inst:24}Centre for Exoplanet Science, SUPA School of Physics and Astronomy, University of St Andrews, North Haugh, St Andrews KY16 9SS, UK \and
\label{inst:25}ESTEC, European Space Agency, 2201AZ, Noordwijk, NL \and
\label{inst:26}Centre for Mathematical Sciences, Lund University, Box 118, 221 00 Lund, Sweden \and
\label{inst:27}Leiden Observatory, University of Leiden, PO Box 9513, 2300 RA Leiden, The Netherlands \and
\label{inst:28}Department of Space, Earth and Environment, Chalmers University of Technology, Onsala Space Observatory, 43992 Onsala, Sweden \and
\label{inst:29}Dipartimento di Fisica, Universita degli Studi di Torino, via Pietro Giuria 1, I-10125, Torino, Italy \and
\label{inst:30}University of Vienna, Department of Astrophysics, Türkenschanzstrasse 17, 1180 Vienna, Austria \and
\label{inst:31}Science and Operations Department - Science Division (SCI-SC), Directorate of Science, European Space Agency (ESA), European Space Research and Technology Centre (ESTEC),Keplerlaan 1, 2201-AZ Noordwijk, The Netherlands \and
\label{inst:32}Konkoly Observatory, Research Centre for Astronomy and Earth Sciences, 1121 Budapest, Konkoly Thege Miklós út 15-17, Hungary \and
\label{inst:33}ELTE Eötvös Lor\'and University, Institute of Physics, P\'azm\'any P\'eter s\'et\'any 1/A, 1117 Budapest, Hungary \and
\label{inst:34}IMCCE, UMR8028 CNRS, Observatoire de Paris, PSL Univ., Sorbonne Univ., 77 av. Denfert-Rochereau, 75014 Paris, France \and
\label{inst:35}Institut d'astrophysique de Paris, UMR7095 CNRS, Université Pierre \& Marie Curie, 98bis blvd. Arago, 75014 Paris, France \and
\label{inst:36}Department of Astrophysics, University of Vienna, Tuerkenschanzstrasse 17, 1180 Vienna, Austria \and
\label{inst:37}INAF, Osservatorio Astrofisico di Catania, Via S. Sofia 78, 95123 Catania, Italy \and
\label{inst:38}Institute of Optical Sensor Systems, German Aerospace Center (DLR), Rutherfordstrasse 2, 12489 Berlin, Germany \and
\label{inst:39}Dipartimento di Fisica e Astronomia "Galileo Galilei", Universita degli Studi di Padova, Vicolo dell'Osservatorio 3, 35122 Padova, Italy \and
\label{inst:40}Cavendish Laboratory, JJ Thomson Avenue, Cambridge CB3 0HE, UK \and
\label{inst:41}ETH Zürich, Department of Physics, Wolfgang-Pauli-Strasse 27, 8093 Zürich, Switzerland \and
\label{inst:42}Center for Astronomy and Astrophysics, Technical University Berlin, Hardenberstrasse 36, 10623 Berlin, Germany \and
\label{inst:43}Institut für Geologische Wissenschaften, Freie Universität Berlin, 12249 Berlin, Germany \and
\label{inst:44}ELTE Eötvös Loránd University, Gothard Astrophysical Observatory, 9700 Szombathely, Szent Imre h. u. 112, Hungary \and
\label{inst:45}MTA-ELTE Exoplanet Research Group, 9700 Szombathely, Szent Imre h. u. 112, Hungary \and
\label{inst:46}MTA-ELTE Lendület ``Momentum'' Milky Way Research Group, Hungary \and
\label{inst:47}Institute of Astronomy, University of Cambridge, Madingley Road, Cambridge, CB3 0HA, United Kingdom
}


\date{Received date / Accepted date }

\abstract{We report the detection of the secondary eclipse of the hot Jupiter HD\,209458\,b in optical/visible light using the CHEOPS space telescope.  Our measurement of \Mdepth\ parts per million translates into a geometric albedo of $A_g =$ \Ag. The previously estimated dayside temperature of about 1500\,K implies that our geometric albedo measurement consists predominantly of reflected starlight and is largely uncontaminated by thermal emission.
This makes the present result one of the most robust measurements of $A_g$ for any exoplanet. 
Our calculations of the bandpass-integrated geometric albedo demonstrate that the measured value of $A_g$ is consistent with a cloud-free atmosphere, where starlight is reflected via Rayleigh scattering by hydrogen molecules, and the water and sodium abundances are consistent with stellar metallicity. We predict that the bandpass-integrated TESS geometric albedo is too faint to detect and that a phase curve of HD\,209458\,b observed by CHEOPS would have a distinct shape associated with Rayleigh scattering if the atmosphere is indeed cloud free. }


\maketitle

\section{Introduction}

The albedo of an exoplanet determines how much starlight enters its atmosphere and hence its global energy budget.  Measurements of the secondary eclipse (occultation) depth of a transiting exoplanet directly yield the albedo at full phase \citep{sea10}, which is known as the geometric albedo \citep{rus16}, if only reflected starlight is measured \citep{hen13} or if thermal emission is accounted for using complementary infrared data \citep{won20,won21}. Such measurements have been made using the Kepler \citep{hen13, ang15, est15}, TESS \citep{won20,won21}, Hubble \citep{swa09,bea17,eva13}, and CHEOPS \citep{len20,hoo21}  space telescopes.  The most convincing measurement of the geometric albedo of an exoplanet is that of the hot Jupiter Kepler-7b using four years of Kepler data \citep{kip11,bri11,bri13,ang15}, although its exact value has been debated in the literature \citep{est15, hen21}.

HD\,209458\,b has been and remains one of the most frequently studied hot Jupiters in history since its discovery in 1999 \citep{hen00,cha00}.  Its Spitzer 4.5\,$\mu$m phase curve reveals a dayside brightness temperature of about 1500\,K \citep{zel14,eva15}. Its optical/visible transmission spectrum, measured by the STIS instrument on board the Hubble Space Telescope (HST), indicates a spectral slope at wavelengths $\lesssim 600$\,nm due to Rayleigh scattering \citep{lec08,sin16}.  Its dayside emission spectrum suggests a water abundance consistent with stellar metallicity and a cloud-free atmosphere at the wavelengths probed by HST-WFC3 \citep{lin16}. Observations of Fe+ and Mg \citep{cub20} might suggest clouds along the terminator \citep{gao20}, which would contradict observations of HST-WFC3 versus J bands in transmission that indicate a cloud-free terminator \citep{ste16}.  
A potential source of opacity in a cloud-free atmosphere is neutral sodium through its strong resonance lines, whose line wings dominate over a very broad spectral range \citep{sud00}.
Na has been observed in transmission \citep{cha02,sin08,sne08,jen11}, although these results have recently been suggested to be analysis artefacts \citep{cas20,cas21}. 
Na and/or TiO in the atmosphere are found to explain the broadband transmission spectrum of the planet best \citep{san20}. Na has never been observed in emission for HD\,209458\,b (D.K.\ Sing, 2021, private communication). Other elements of interest are atomic hydrogen, oxygen, and carbon, which have been reported in the extended escaping atmosphere of the planet \citep{vid03,vid04}.

Previous efforts to detect the optical/visible occultation of HD\,209458\,b using the MOST space telescope yielded upper limits \citep[$A_g < 0.17\,{[3\sigma]}$ in the MOST bandpass,][]{row06,row08}. In this Letter, we report a robust $>6\sigma$ detection of the occultation depth of HD\,209458\,b using the CHEOPS space telescope \citep{ben21} and demonstrate that the corresponding geometric albedo is consistent with a cloud-free atmosphere of a chemical abundance similar to that of the star.



\section{Methods}

\subsection{Observations and data processing\label{s:obs}}
We observed ten occultations of HD\,209458\,b between July and September 2021 (CHEOPS programme CH\_PR100016; see observation log in Table\,\ref{tab:cheops_log}). Each visit lasted for about 11\,h, corresponding to seven CHEOPS orbits with an efficiency (fraction of time on target) of $\sim$70\%. To save bandwidth, groups of three 11.5\,s exposures of a circular subarray with a radius of 100 pixels were co-added prior to downlink, resulting in a cadence of 34.5\,s. In addition, single-exposure circular \textit{\textup{imagettes}} with a radius of 30 pixels were downloaded. We analysed the subarrays using photometry from the \textit{Data Reduction Pipeline} \citep[\texttt{DRP}; ][]{hoy20} and the point-spread function photometry package \texttt{PIPE}\footnote{\url{https://github.com/alphapsa/PIPE}} 
developed specifically for CHEOPS (Brandeker et al.\ in prep.; see also descriptions in \citealt{sza21} and \citealt{mor21}) as well as imagette photometry with \texttt{PIPE}.
We found that the results were consistent with a $\sim$100\,ppm in standard deviation over 1\,min bins (for this G = 7.5\,mag star), but the imagette photometry resulted in a scatter lower by 10\% in the measured occultation depths. This is likely due to the faster cadence of the imagettes, which allows detrending with a higher time-resolution. We thus here focus on the analysis of the imagette photometry.

The light curves were analysed using the \texttt{pycheops} software, a Python module specifically developed to analyse light curves from CHEOPS \citep{max21}. Data points marked as poor by \texttt{PIPE} (due to e.g. strong cosmic rays or contamination from a satellite passing through the field of view; this affects 527 out of 25\,866, i.e.,  2.0\% of the data points) were masked from the analysis. We also removed data with a background higher than 300\,e$^-$\,pix$^{-1}$ (an additional 652 out of 25\,866, i.e.,  2.5\%) because they are empirically difficult to reduce adequately, possibly due to non-linear sensitivity linked to charged-transfer inefficiency. The poor data points were flagged at the data reduction stage, uninformed by the extracted light curve. The end result are therefore not biased by the filtering.
Each visit was analysed individually, for which we fixed the transit depth and used informative priors based on their literature values given to the transit width, impact parameter, and transit central time (Table~\ref{tab:params}). The orbit was assumed to be circular \citep{dem05,cro12}, such that the occultation time is shifted by half an orbit compared to the transit, corrected for light-travel time.  The occultation depth was generally given a broad uniform prior (1--200\,ppm), with the exception for visit 5, where the Monte Carlo Markov chain (MCMC) analysis did not converge (using $10^3$--$10^4$ steps and 128--512 walkers). Instead, the occultation depth for visit 5 was derived with an unconstraining improper prior $\mathcal{U}\!\left(-\infty, \infty\right)$, resulting in a negative occultation depth (Table~\ref{tab:eclipses}). It is statistically expected for single visits that a negative occultation depth may be derived because the signal-to-noise ratio per visit is low. An initial fit of individual visits was made using the least-square minimisation Python package \texttt{lmfit,} which includes roll angle and (x,~y)-position decorrelation (selected so that it results in a Bayes factor $<1$). The residuals were then fitted by a Gaussian process (GP), which give strong priors on the GP (based on a preliminary fit in which only GP parameters were free) and the planetary parameters (from the \texttt{lmfit} analysis) for the subsequent GP fit. In the end, all of these visits were analysed simultaneously using the MCMC scheme with the \texttt{MultiVisit} method within \texttt{pycheops,} in which the priors were inferred from the fit to the individual visits.
%
To search for remaining correlations in the residuals, we used the Shapiro-Wilk test \citep{sha65}.  The resulting p-value of 0.72 indicates that the residuals are indistinguishable from white noise.

To convert the measured occultation depth into a geometric albedo, we first estimated the contribution from thermal emission by extrapolating the daytime temperature as found by \textit{Spitzer} 4.5\,$\mu$m occultation measurements \citep{zel14} to the CHEOPS bandpass assuming an irradiated model atmosphere from \citet{mol15} with the parameters $T_{\mathrm{eff}}=1500$\,K, [Fe/H] = 0.0, C/O = 0.55, irradiating spectral type F5, and $\log g = 3.0$. We then computed the geometric albedo as $A_g = (a/R_p)^2 L$, where $L$ is the occultation depth corrected for thermal emission, and the ratio $a/R_p$ was computed from $a/R_\star$ and $R_p/R_\star$ found in Table~\ref{tab:params}.

We estimated the TESS expected detection threshold for the HD\,209458\,b occultation depth by assuming the
sensitivity to be equivalent to existing TESS observations of the similar system HD\,189733\,b in sector 41. We also computed the thermal contribution in the TESS and MOST bands using the same model atmosphere as before.

\begin{table*}
    \centering
    \caption{Logs of CHEOPS observations.}
    \begin{tabular}{cccccc}
    \hline
    \hline
    \noalign{\smallskip}
    Visit & Start Date & End Date & File Key & Num.\ of & Num.\ of     \\
     \# & (2021) & (2021) & & subarrays & imagettes \\
    \noalign{\smallskip}
    \hline
    \noalign{\smallskip}
1 & {\small 2021-07-27 11:09:49} & {\small 2021-07-27 22:09:36} & {\small PR100016\_TG013701} & 830 & 2490 \\
2 & {\small 2021-08-03 12:11:49} & {\small 2021-08-03 23:21:58} & {\small PR100016\_TG013702} & 861 & 2583 \\
3 & {\small 2021-08-06 23:50:49} & {\small 2021-08-07 11:59:10} & {\small PR100016\_TG013703} & 911 & 2733 \\
4 & {\small 2021-08-10 12:25:49} & {\small 2021-08-10 23:35:58} & {\small PR100016\_TG013704} & 877 & 2631 \\
5 & {\small 2021-08-21 03:09:49} & {\small 2021-08-21 14:11:55} & {\small PR100016\_TG013705} & 901 & 2703 \\
6 & {\small 2021-08-31 17:06:49} & {\small 2021-09-01 04:07:45} & {\small PR100016\_TG013706} & 884 & 2652 \\
7 & {\small 2021-09-04 04:34:49} & {\small 2021-09-04 15:44:58} & {\small PR100016\_TG013707} & 876 & 2628 \\
8 & {\small 2021-09-11 05:45:48} & {\small 2021-09-11 17:06:55} & {\small PR100016\_TG013708} & 886 & 2658 \\
9 & {\small 2021-09-18 07:03:59} & {\small 2021-09-18 18:06:40} & {\small PR100016\_TG013709} & 840 & 2520 \\
10 & {\small 2021-09-21 20:58:49} & {\small 2021-09-22 07:47:39} & {\small PR100016\_TG013710} & 756 & 2268 \\
    \noalign{\smallskip}
  \hline
    \end{tabular}
    \tablefoot{Time notation follows the ISO-8601 convention. The File Key aids the retrieval of data from the CHEOPS archive.}
    \label{tab:cheops_log}
\end{table*}

\begin{table}
    \centering
    \caption{Measured occultation depths.}
    \begin{tabular}{cccc}
    \hline
    \hline
    \noalign{\smallskip}
    Visit & Mid occultation & Depth & $\sigma$ \\
     \# & (BJD$-$2\,459\,400) & [ppm] & [ppm] \\
    \noalign{\smallskip}
    \hline
    \noalign{\smallskip}
1 & 23.20009 & 14.8 & 11.6 \\
2 & 30.24419 & 31.9 & 9.3 \\
3 & 33.77430 & 25.2 & 9.8 \\
4 & 37.29726 & 10.3 & 8.3 \\
5 & 47.86952 & -1.6 & 10.8 \\
6 & 58.44554 & 21.8 & 10.1 \\
7 & 61.97386 & 31.0 & 9.5 \\
8 & 69.01787 & 24.5 & 9.9 \\
9 & 76.06736 & 40.4 & 10.0 \\
10 & 79.59218 & 17.1 & 13.2 \\
    \noalign{\smallskip}
  \hline
    \end{tabular}
    \label{tab:eclipses}
\end{table}

\begin{figure}
\includegraphics[width=\columnwidth]{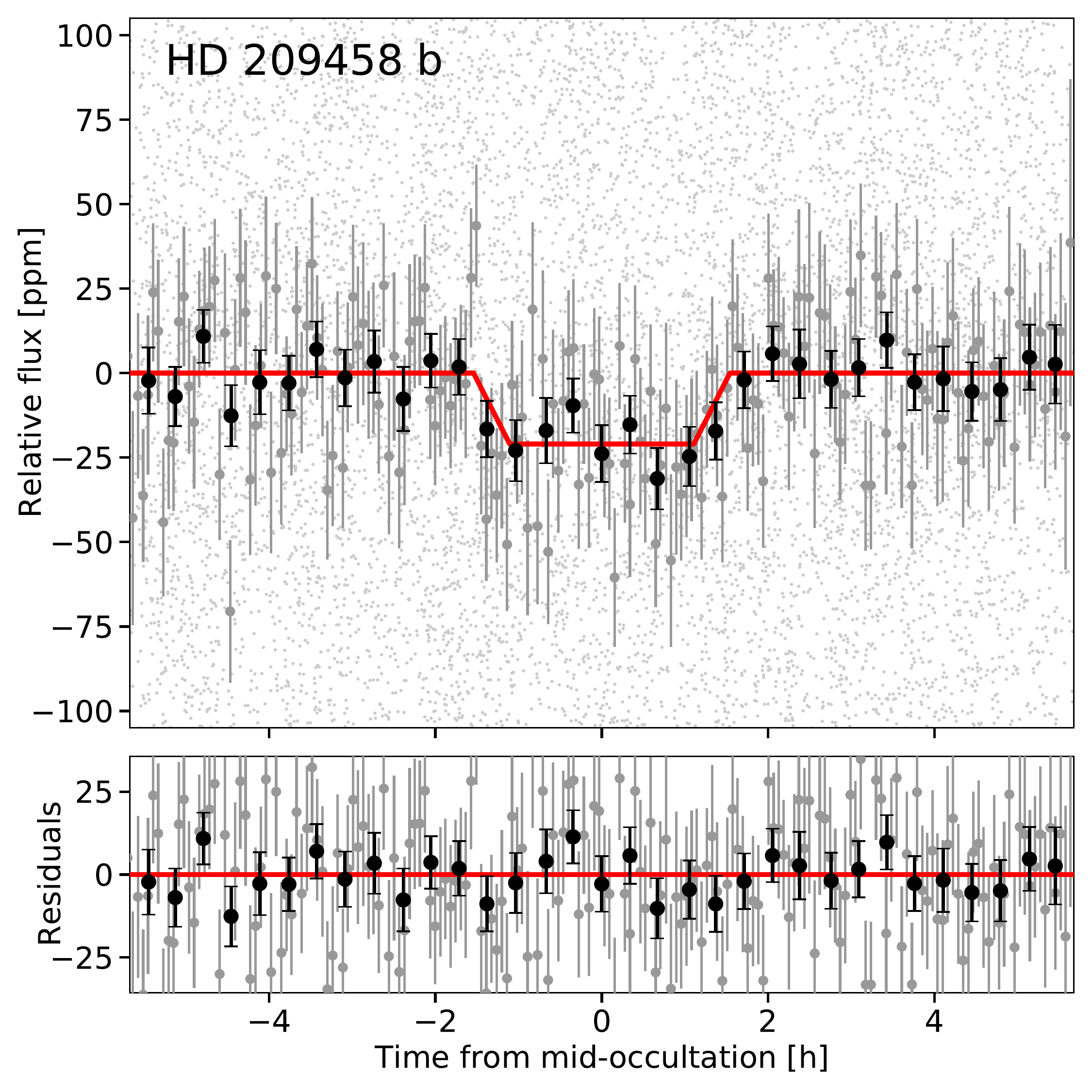}
\vspace{-0.2in}
\caption{ Phase-folded data with best-fit occultation model (unbroken red line). Individual imagette measurements have an average photometric uncertainty of 237\,ppm and fill the figure with small, light grey points. To better visualise the occultation, data points are also shown averaged to 3.7\,min bins (grey) and 20.5\,min bins (black). The bin sizes are chosen to divide the occultation duration into an integer number of bins. The lower panel shows the residuals after the occultation model was subtracted from the data.}
\label{fig:phasefolded}
\end{figure}

\begin{figure}
\includegraphics[width=\columnwidth]{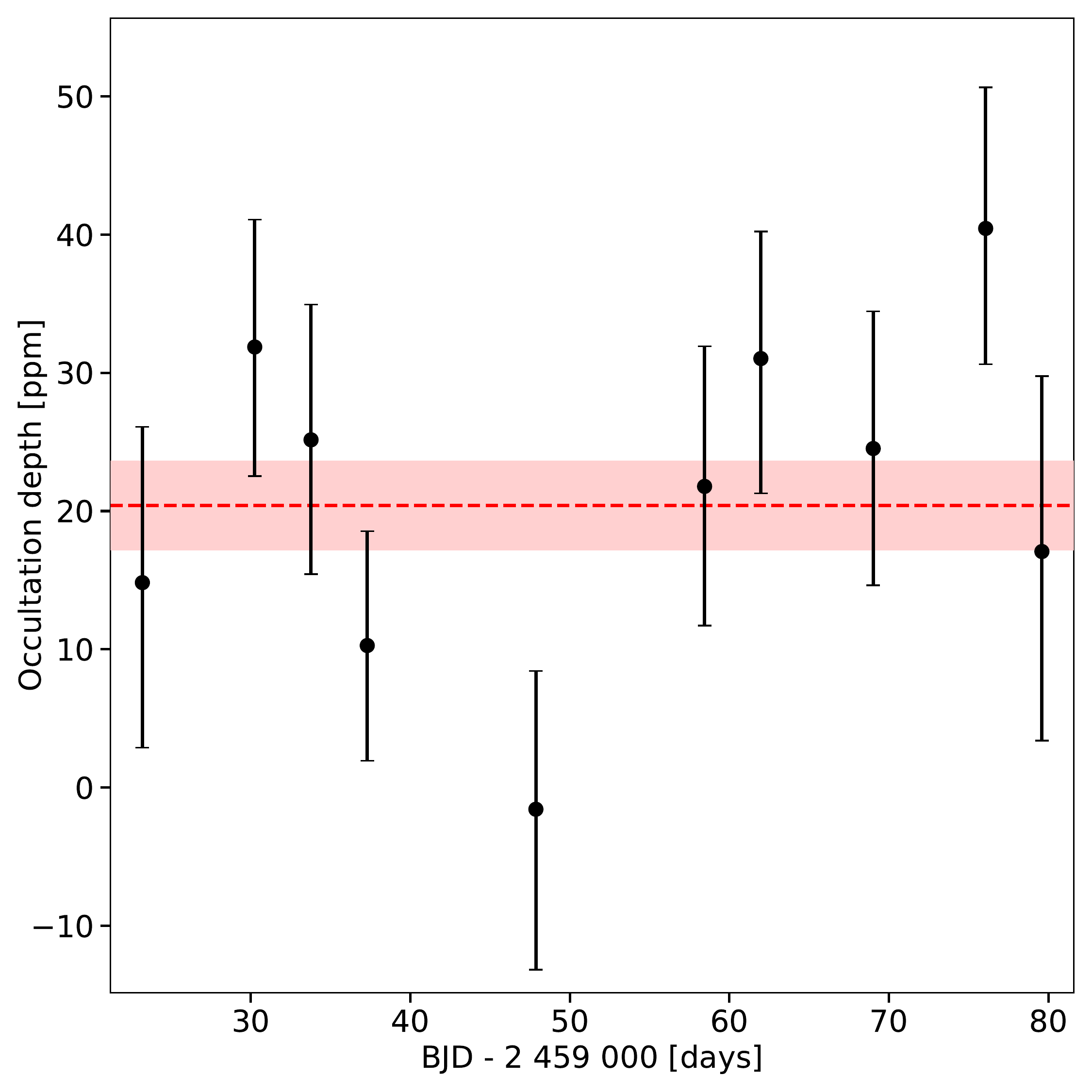}
\vspace{-0.2in}
\caption{Occultation measurements from the ten individual visits, plotted against the time of their observation (in barycentric Julian dates, BJD). The dashed red line is the multivisit-fitted depth \Mdepth\,ppm, and the shaded red region shows the 1$\sigma$ uncertainty region.}
\label{fig:individual}
\end{figure}

\subsection{Calculations of geometric albedos}

\begin{figure*}
\begin{center}
\vspace{-0.2in} 
\includegraphics[width=\columnwidth]{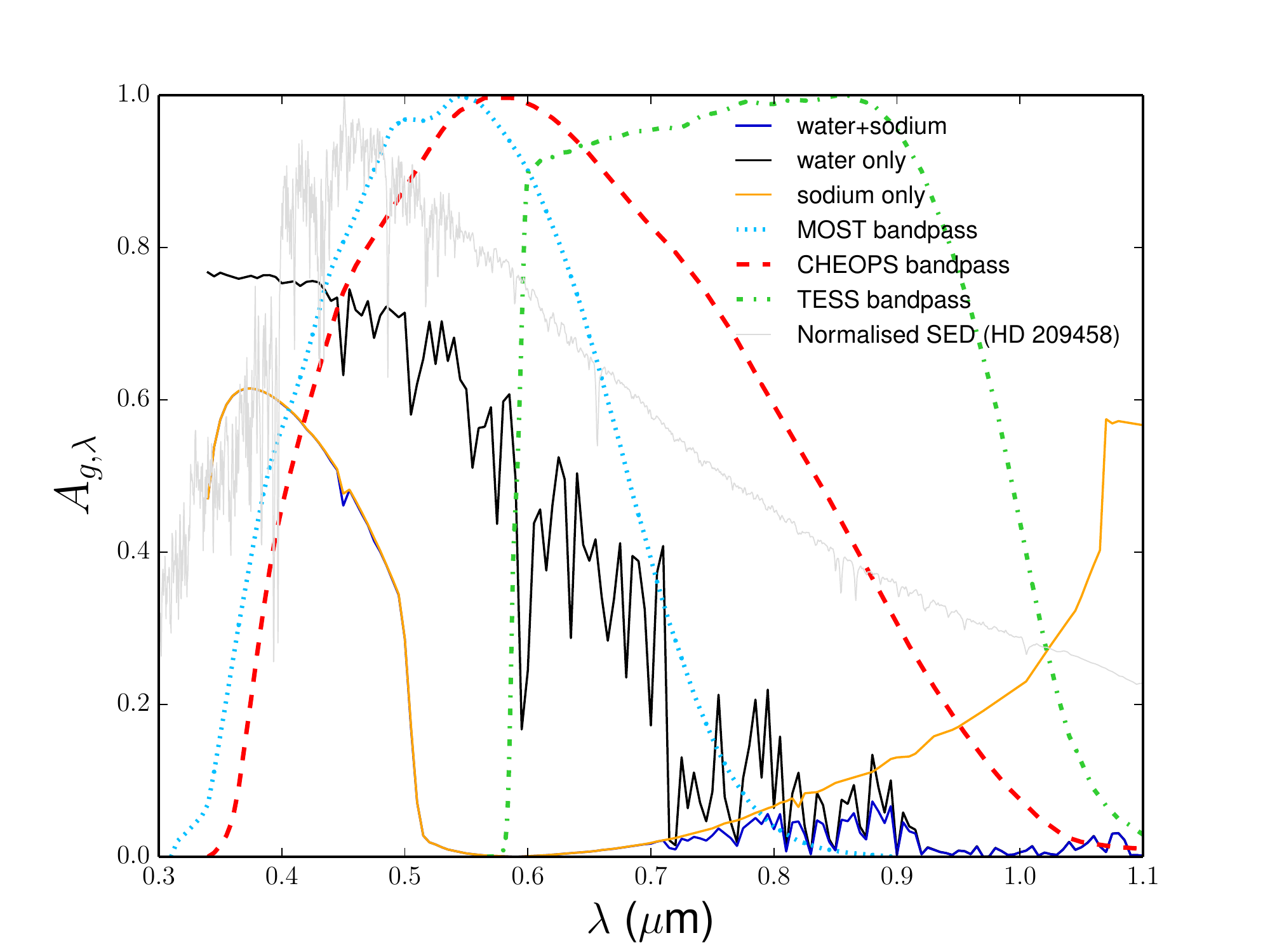}
\includegraphics[width=\columnwidth]{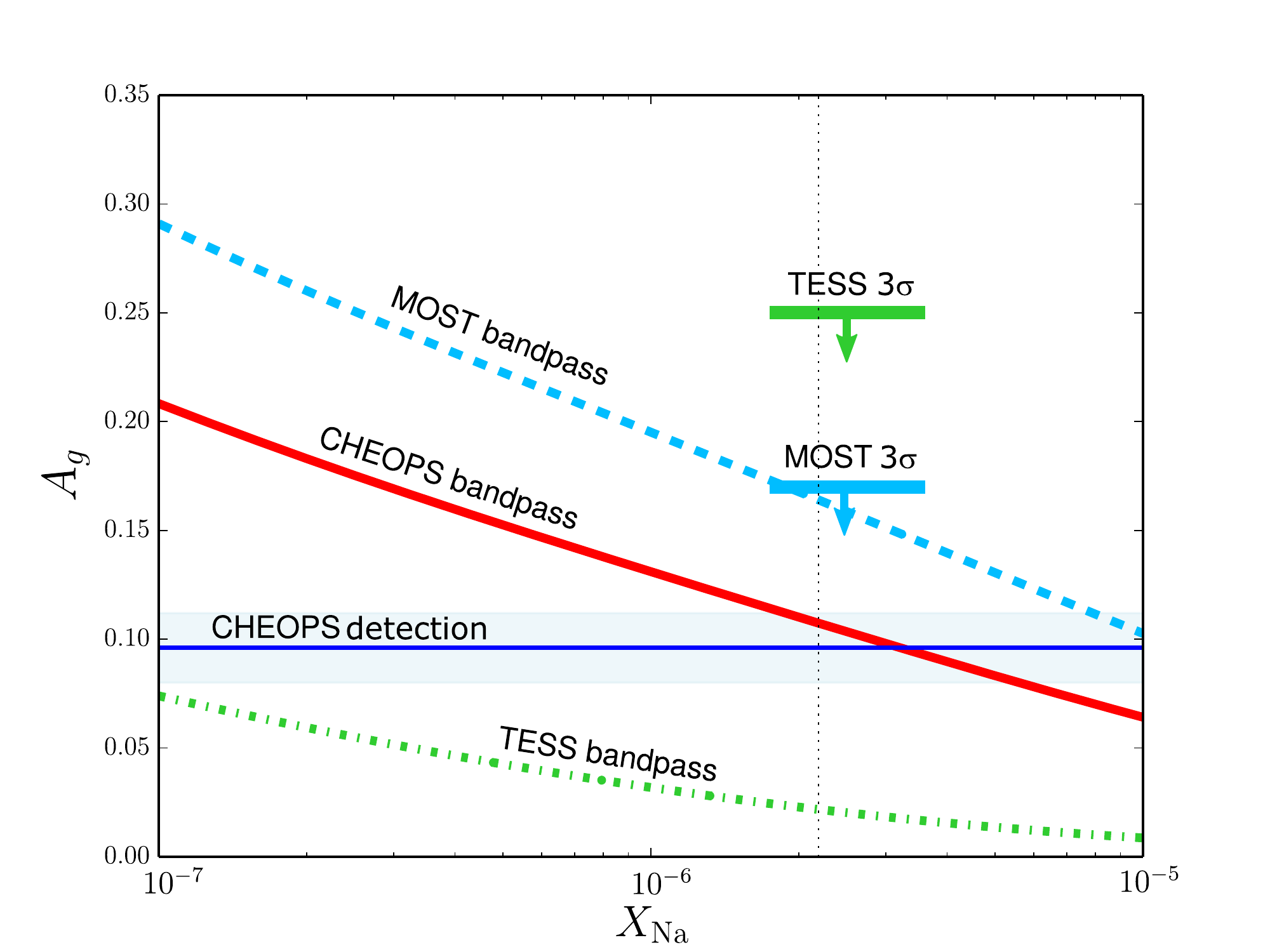}
\end{center}
\vspace{-0.2in}
\caption{Bandpass-dependent geometric albedos. Left panel: Examples of geometric albedo spectra assuming stellar abundances of H$_2$O ($X_{\rm H_2O}=5.36 \times 10^{-4}$) and/or Na ($X_{\rm Na}=2.19 \times 10^{-6}$).  The corresponding bandpass-integrated CHEOPS geometric albedos for H$_2$O and Na, H$_2$O only, and Na only are $A_g = 0.11$, 0.41, and 0.12, respectively.  The bandpass filters of the CHEOPS, MOST, and TESS space telescopes are overlaid, as are the normalised spectral energy distribution of the HD\,209458 star (where the fluxes have been divided by their maximum value such that they have no physical units).  Right panel: Modelled bandpass-integrated geometric albedos for MOST, CHEOPS, and TESS bandpasses assuming a stellar abundance of H$_2$O, but with varying volume mixing ratios of Na. The vertical dotted line corresponds to $X_{\rm Na}=2.19 \times 10^{-6}$, the Na abundance of the star.  The horizontal line with the associated shaded region depicts the $A_g =$ \Ag\ measurement reported in this study, and the MOST and predicted TESS 3$\sigma$ upper limits are added as horizontal lines. All calculations assume a temperature of 1500\,K; calculations with 1400\,K produce a very similar outcome (not shown).}
\label{fig:albedo}
\end{figure*}

The monochromatic geometric albedo associated with Rayleigh scattering is \citep{hen21}
\begin{equation}
A_{g,\lambda} = \frac{\omega}{16} + \frac{\epsilon}{2} + \frac{\epsilon^2}{6} + \frac{\epsilon^3}{24},
\end{equation}
where $\omega$ is the single-scattering albedo, $\epsilon = (1-\gamma)/(1+\gamma)$ is the bihemispherical reflectance \citep{hap81}, and $\gamma = \sqrt{1-\omega}$.  In the preceding expression, the first term accounts for single scattering of starlight, while the other terms account for multiple scattering.  The single-scattering albedo is given by the ratio of cross sections,
\begin{equation}
\omega = \frac{0.9\sigma_{\rm H_2}}{0.9\sigma_{\rm H_2} + \sigma_{\rm abs}},
\end{equation}
where the Rayleigh scattering cross section due to molecular hydrogen is \citep{cox00}
\begin{equation}
\sigma_{\rm H_2} = \frac{24 \pi^3}{n^2_{\rm ref} \lambda^4} \left( \frac{k^2-1}{k^2+2} \right)^2,
\end{equation}
$n_{\rm ref} = 2.68678 \times 10^{19}$ cm$^{-3}$, $\lambda$ is the wavelength, and the refractive index is \begin{equation}
k = 1.358 \times 10^{-4} \left[ 1 + 0.00752 \left(\frac{\lambda}{1 
~\mu\mbox{m}} \right)^{-2} \right] + 1.
\end{equation}
Commensurate with the limited amount of information that is measured, the absorption cross section is assumed to depend only on H$_2$O and Na,
\begin{equation}
\sigma_{\rm abs} = X_{\rm H_2O} ~\sigma_{\rm H_2O} + X_{\rm Na} ~\sigma_{\rm Na},
\end{equation}
where the mixing ratios of H$_2$O and Na are denoted by $X_{\rm H_2O}$ and $X_{\rm Na}$, respectively. Other absorbers such as CH$_4$ or CO are not sufficiently abundant to produce significant opacities in the considered bandpass.

The cross section of H$_2$O ($\sigma_{\rm H_2O}$) was computed using input from the \texttt{ExoMol} spectroscopic line-list database \citep{pol18} and using the open-source \texttt{HELIOS-K} opacity calculator\footnote{\url{https://github.com/exoclime/HELIOS-K}} \citep{gri15,gri21}. It is publicly available in the Swiss-based DACE opacity database\footnote{\url{https://dace.unige.ch}} \citep{gri21}.  

The cross section of Na ($\sigma_{\rm Na}$) was computed using Kurucz line-list data \citep{kur95}. We accounted for the natural line width, thermal broadening, and pressure broadening using the tabulated van der Waals broadening coefficients from the line list. The latter ones were scaled from collisions with atomic hydrogen to those of collisions with molecular hydrogen by using the Uns\"{o}ld approximation \citep{uns55}. The strong resonance lines of Na are known to deviate from the usual Voigt profiles. Their strongly non-Lorentzian far-wing line profiles originating from collisions with H$_2$ were calculated based on \citet{all19}.

To compute these cross sections, we assumed a temperature of 1500\,K. This is consistent with the brightness temperature measured by Spitzer \citep{zel14,eva15}. We also assumed a pressure of 0.1\,bar \citep{lin16}.

We computed $X_{\rm H_2O}$ using previously benchmarked formulae \citep{hen16}, assuming the chemical abundance of the Sun \citep{lod03} scaled to that of the star \citep[close to the solar value at {[Fe/H]} = $0.04 \pm 0.01$; ][]{sou21}. This yielded the input elemental abundances of carbon, oxygen, and nitrogen: $\mbox{C/H}=2.69 \times 10^{-4}$, $\mbox{O/H}=5.37 \times 10^{-4}$, and $\mbox{N/H}=7.41 \times 10^{-5}$.  These elemental abundances correspond to $\mbox{C/O}=0.50$ and $\mbox{N/O}=0.14$.  With these inputs, the volume mixing ratio of H$_2$O is $X_{\rm H_2O}=5.36 \times 10^{-4}$.

Formally, $A_{g,\lambda}$ is defined at a single wavelength.  In practice, the occultation depth is measured across a range of optical/visible wavelengths, which yields a bandpass-integrated geometric albedo,
\begin{equation}
A_g = \frac{\int A_{g,\lambda} ~\lambda ~F_\star ~{\cal F} ~d\lambda}{\int \lambda ~F_\star ~{\cal F} ~d\lambda},
\end{equation}
where $F_\star$ is the spectral energy distribution of the star, and ${\cal F}$ is the bandpass filter.  For the HD\,209458 star, we obtained $F_\star$ from the \textit{HST} Calibration Database CALSPEC\footnote{The FITS file \texttt{hd209458\_stisnic\_006.fits} available at \url{http://www.stsci.edu/hst/observatory/crds/calspec.html}} \citep{boh14}.  The bandpass filters for the CHEOPS, MOST, and TESS telescopes were downloaded from the SVO filter service \citep{rod12,rod20}.  The factor of $\lambda$ in the integrands accounts for the conversion of energy into photon flux.

\section{Results}

From the simultaneous fit to all visits, we find the occultation depth to be $L =$ \Mdepth\,ppm (Fig.~\ref{fig:phasefolded}). This is consistent with the average over individual fits (Fig.~\ref{fig:individual}, Table~\ref{tab:eclipses}), $L =$ \Adepth\,ppm. Adopting the simultaneous fit and subtracting a 2.2\,ppm contribution from thermal emission, we find a geometric albedo $A_g =$ \Ag.

In the left panel of Fig.~\ref{fig:albedo}, we show examples of geometric albedo spectra $A_{g,\lambda}$, derived assuming an atmosphere of the same chemical composition as the star (with $X_{\rm Na}=2.19 \times 10^{-6}$ and $X_{\rm H_2O}=5.36 \times 10^{-4}$ from chemical balance).  If H$_2$O is the only absorber present, then we obtain a CHEOPS bandpass-integrated geometric albedo of $A_g = 0.41$, which is much higher than the measurement.  If Na and H$_2$O are included, then we obtain $A_g = 0.11$; if only Na is considered, we obtain $A_g = 0.12$. This confirms the prediction of \citet{sud00}: one reason why hot Jupiters are dark in the optical/visible range of wavelengths is absorption by the resonant doublet of Na \citep[see also][]{sea00}.  

In theory, the geometric albedo is a monochromatic quantity.  In practice, the bandpass-integrated geometric albedo $A_g$ differs when it is measured by the MOST, CHEOPS, and TESS space telescopes because the bandpass filters cover different ranges of wavelengths (left panel of Fig.~\ref{fig:albedo}).  Our calculations indicate that Na is the dominant absorber in shaping the albedo spectrum across optical/visible wavelengths \citep{sea00,sud00}.  In the right panel of Fig.~\ref{fig:albedo}, we compute $A_g$ for all three bandpasses and allow the relative abundance (by number) of Na to vary.  The measured CHEOPS geometric albedo of \Ag\ is consistent with the theoretical geometric albedo derived from a stellar Na abundance. This does not rule out the possibility that other relative abundances of elements may reproduce $A_g$, but the close match between the measured and predicted $A_g$ from the simplest hypothesis of an atmosphere with the same chemical abundance as the star is remarkable.

\section{Conclusions and discussion}

The right panel of Fig.~\ref{fig:albedo} demonstrates that the TESS geometric albedo is expected to be $\lesssim 0.03$ for $X_{\rm Na} \sim 10^{-6}$. This translates into an occultation depth of $\lesssim 6$\,ppm in the TESS band from purely reflected light. When we add a 8.5\,ppm thermal contribution as estimated from the model atmosphere (Sect.~\ref{s:obs}), we obtain a total expected occultation depth of $\lesssim 15$\,ppm. This is not detectable by TESS because we estimate that its 3$\sigma$ detection threshold would be $\sim 50$\,ppm for an observation duration of a single sector (corresponding to $A_g = 0.25$).

\begin{figure}
\includegraphics[width=\columnwidth]{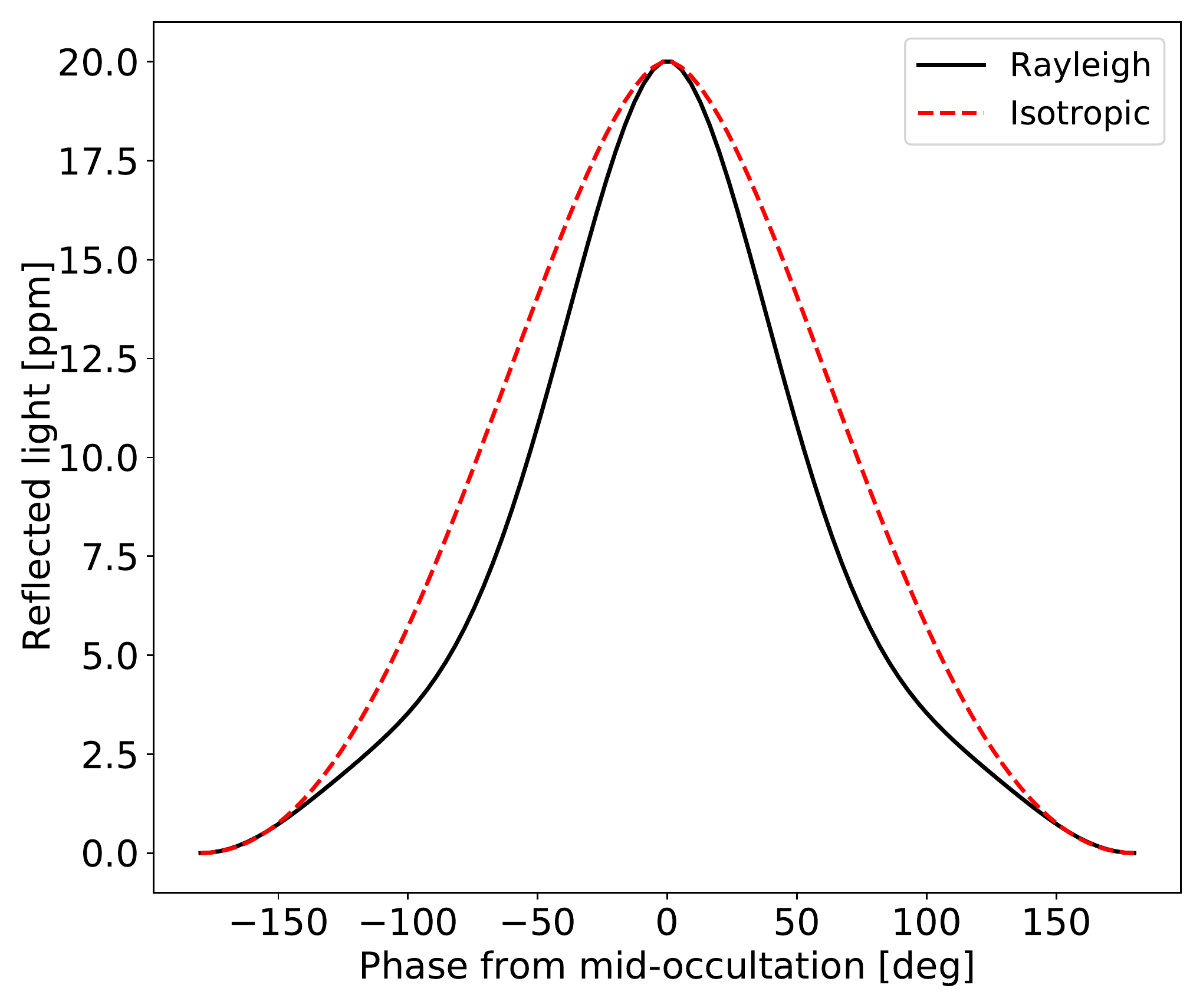}
\vspace{-0.2in}
\caption{Predicted reflected light relative to the star as a function of phase for Rayleigh and isotropic scattering in the CHEOPS photometric band from HD\,209458\,b, computed using the formulation in \citet{hen21}. The single-scattering albedos $\omega$  were 0.45 for Rayleigh scattering and 0.55  for isotropic scattering.
While the phase curve itself should be detectable with CHEOPS, it will be challenging to distinguish between the two scattering laws. }
\label{fig:phasecurves}
\end{figure}

As already pointed out by \citet{hen21}, reflected-light phase curves that are free of clouds are expected to have a distinct shape associated with Rayleigh scattering (Fig.~\ref{fig:phasecurves}). Using Monte Carlo simulations of observations, we find that the noise must be lower than 4.5\,ppm per hour of the phase curve in order to distinguish Rayleigh from isotropic scattering with better than 95\% confidence. This will be challenging to achieve considering the variability of the star \citep{row06,row08}. Fortunately, confusion of the orbital phase with stellar rotation modulation will be mitigated by their very different timescales
\citep[the rotation period is $\text{about two}$\,weeks, as inferred from][]{cas21}.


The measured geometric albedo of HD\,209458\,b is consistent with a cloud-free, stellar-metallicity atmosphere. This corroborates the findings of \citet{sud00}, who predicted that ``Class IV roasters'', as they called them, have low geometric albedos due to absorption by alkali metals \citep[see also][]{sea00}.

Our result demonstrates the unique capability of CHEOPS to measure the geometric albedo through ultra-high-precision photometry of occultations at optical wavelengths. It is highly desirable to apply this technique to a larger sample to improve our understanding of planetary atmospheres and their energy balance. A publication is in preparation by the CHEOPS consortium to present updated and more detailed performance estimates for various scenarios, derived from experience with the first two years of data.

\begin{acknowledgements}
CHEOPS  is  an  ESA  mission  in  partnership  with  Switzerland  with  important  contributions  to 
the payload and the ground segment from Austria, Belgium, France, Germany, Hungary, Italy, 
Portugal, Spain, Sweden, and the United Kingdom. The CHEOPS Consortium would like to 
gratefully  acknowledge  the  support  received  by  all  the  agencies,  offices,  universities,  and 
industries involved. Their flexibility and willingness to explore new approaches were essential 
to the success of this mission. 
KGI is the ESA CHEOPS Project Scientist and is responsible for the ESA CHEOPS Guest 
Observers  Programme.  She  does  not  participate  in,  or  contribute  to,  the  definition  of  the 
Guaranteed Time Programme of the CHEOPS mission through which observations described 
in this paper have been taken, nor to any aspect of target selection for the programme.
This research has made use of the SVO Filter Profile Service (http://svo2.cab.inta-csic.es/theory/fps/) supported from the Spanish MINECO through grant AYA2017-84089.
We used matplotlib \citep{hun07} for the figures.
A.Br.\ was supported by the SNSA. 
P.F.L.M.\ acknowledges support from STFC research grant number ST/M001040/1. 
This work was supported by FCT - Fundação para a Ciência e a Tecnologia through national funds and by FEDER through COMPETE2020 - Programa Operacional Competitividade e Internacionalizacão by these grants: UID/FIS/04434/2019, UIDB/04434/2020, UIDP/04434/2020, PTDC/FIS-AST/32113/2017 \& POCI-01-0145-FEDER- 032113, PTDC/FIS-AST/28953/2017 \& POCI-01-0145-FEDER-028953, PTDC/FIS-AST/28987/2017 \& POCI-01-0145-FEDER-028987, O.D.S.D.\ is supported in the form of work contract (DL 57/2016/CP1364/CT0004) funded by national funds through FCT. 
L.D.\ is an F.R.S.-FNRS Postdoctoral Researcher. The Belgian participation to CHEOPS has been supported by the Belgian Federal Science Policy Office (BELSPO) in the framework of the PRODEX Program, and by the University of Liège through an ARC grant for Concerted Research Actions financed by the Wallonia-Brussels Federation. 
B.-O.D.\ acknowledges support from the Swiss National Science Foundation (PP00P2-190080). 
Y.A.\ and M.J.H.\ acknowledge the support of the Swiss National Fund under grant 200020\_172746. 
We acknowledge support from the Spanish Ministry of Science and Innovation and the European Regional Development Fund through grants ESP2016-80435-C2-1-R, ESP2016-80435-C2-2-R, PGC2018-098153-B-C33, PGC2018-098153-B-C31, ESP2017-87676-C5-1-R, MDM-2017-0737 Unidad de Excelencia Maria de Maeztu-Centro de Astrobiolog\'ia (INTA-CSIC), as well as the support of the Generalitat de Catalunya/CERCA programme. The MOC activities have been supported by the ESA contract No. 4000124370. 
S.C.C.B.\ acknowledges support from FCT through FCT contracts nr.\  IF/01312/2014/CP1215/CT0004. 
X.B., S.C., D.G., M.F., and J.L.\ acknowledge their role as ESA-appointed CHEOPS science team members. 
This project was supported by the CNES.~
This project has received funding from the European Research Council (ERC) under the European Union’s Horizon 2020 research and innovation programme (project {\sc Four Aces}.~grant agreement No 724427).~
It has also been carried out in the frame of the National Centre for Competence in Research PlanetS supported by the Swiss National Science Foundation (SNSF). D.E.\ acknowledges financial support from the Swiss National Science Foundation for project 200021\_200726.
M.F.\ was supported by SNSA (DNR 65/19, 174/18). 
D.G.\ was supported by the CRT foundation under Grant No. 2018.2323 ``Gaseous or rocky? Unveiling the nature of small worlds''. 
M.G.\ is an F.R.S.-FNRS Senior Research Associate. 
S.H.\ gratefully acknowledges CNES funding through the grant 837319. 
K.G.I.\ is the ESA CHEOPS Project Scientist and is responsible for the ESA CHEOPS Guest Observers Programme. She does not participate in, or contribute to, the definition of the Guaranteed Time Programme of the CHEOPS mission through which observations described in this paper have been taken, nor to any aspect of target selection for the programme. 
This work was granted access to the HPC resources of MesoPSL financed by the Region Ile de France and the project Equip@Meso (reference ANR-10-EQPX-29-01) of the programme Investissements d'Avenir supervised by the Agence Nationale pour la Recherche.~
This work was partially supported by a grant from the Simons Foundation (PI Queloz, grant number 327127). 
S.G.S.\ acknowledge support from FCT through FCT contract nr. CEECIND/00826/2018 and POPH/FSE (EC).~
This project has been supported by the Hungarian National Research, Development and Innovation Office (NKFIH) grants GINOP-2.3.2-15-2016-00003, K-119517, K-125015, and the City of Szombathely under Agreement No.\ 67.177-21/2016. 
V.V.G.\ is an F.R.S-FNRS Research Associate.


\end{acknowledgements}

\bibliographystyle{aa}
\bibliography{hd209}

\begin{thebibliography}{70}
\expandafter\ifx\csname natexlab\endcsname\relax\def\natexlab#1{#1}\fi

\bibitem[{{Albrecht} {et~al.}(2012){Albrecht}, {Winn}, {Johnson}, {Howard},
  {Marcy}, {Butler}, {Arriagada}, {Crane}, {Shectman}, {Thompson}, {Hirano},
  {Bakos}, \& {Hartman}}]{alb12}
{Albrecht}, S., {Winn}, J.~N., {Johnson}, J.~A., {et~al.} 2012, \apj, 757, 18

\bibitem[{{Allard} {et~al.}(2019){Allard}, {Spiegelman}, {Leininger}, \&
  {Molliere}}]{all19}
{Allard}, N.~F., {Spiegelman}, F., {Leininger}, T., \& {Molliere}, P. 2019,
  \aap, 628, A120

\bibitem[{{Angerhausen} {et~al.}(2015){Angerhausen}, {DeLarme}, \&
  {Morse}}]{ang15}
{Angerhausen}, D., {DeLarme}, E., \& {Morse}, J.~A. 2015, \pasp, 127, 1113

\bibitem[{{Beatty} {et~al.}(2017){Beatty}, {Madhusudhan}, {Tsiaras}, {Zhao},
  {Gilliland}, {Knutson}, {Shporer}, \& {Wright}}]{bea17}
{Beatty}, T.~G., {Madhusudhan}, N., {Tsiaras}, A., {et~al.} 2017, \aj, 154, 158

\bibitem[{{Benz} {et~al.}(2021){Benz}, {Broeg}, {Fortier}, {Rando}, {Beck},
  {Beck}, {Queloz}, {Ehrenreich}, {Maxted}, {Isaak}, {Billot}, {Alibert},
  {Alonso}, {Ant{\'o}nio}, {Asquier}, {Bandy}, {B{\'a}rczy}, {Barrado},
  {Barros}, {Baumjohann}, {Bekkelien}, {Bergomi}, {Biondi}, {Bonfils},
  {Borsato}, {Brandeker}, {Busch}, {Cabrera}, {Cessa}, {Charnoz}, {Chazelas},
  {Collier Cameron}, {Corral Van Damme}, {Cortes}, {Davies}, {Deleuil},
  {Deline}, {Delrez}, {Demangeon}, {Demory}, {Erikson}, {Farinato}, {Fossati},
  {Fridlund}, {Futyan}, {Gandolfi}, {Garcia Munoz}, {Gillon}, {Guterman},
  {Gutierrez}, {Hasiba}, {Heng}, {Hernandez}, {Hoyer}, {Kiss}, {Kovacs},
  {Kuntzer}, {Laskar}, {Lecavelier des Etangs}, {Lendl}, {L{\'o}pez}, {Lora},
  {Lovis}, {L{\"u}ftinger}, {Magrin}, {Malvasio}, {Marafatto}, {Michaelis}, {de
  Miguel}, {Modrego}, {Munari}, {Nascimbeni}, {Olofsson}, {Ottacher},
  {Ottensamer}, {Pagano}, {Palacios}, {Pall{\'e}}, {Peter}, {Piazza}, {Piotto},
  {Pizarro}, {Pollaco}, {Ragazzoni}, {Ratti}, {Rauer}, {Ribas}, {Rieder},
  {Rohlfs}, {Safa}, {Salatti}, {Santos}, {Scandariato}, {S{\'e}gransan},
  {Simon}, {Smith}, {Sordet}, {Sousa}, {Steller}, {Szab{\'o}}, {Szoke},
  {Thomas}, {Tschentscher}, {Udry}, {Van Grootel}, {Viotto}, {Walter},
  {Walton}, {Wildi}, \& {Wolter}}]{ben21}
{Benz}, W., {Broeg}, C., {Fortier}, A., {et~al.} 2021, Experimental Astronomy,
  51, 109

\bibitem[{{Bohlin} {et~al.}(2014){Bohlin}, {Gordon}, \& {Tremblay}}]{boh14}
{Bohlin}, R.~C., {Gordon}, K.~D., \& {Tremblay}, P.~E. 2014, \pasp, 126, 711

\bibitem[{{Bonomo} {et~al.}(2017){Bonomo}, {Desidera}, {Benatti}, {Borsa},
  {Crespi}, {Damasso}, {Lanza}, {Sozzetti}, {Lodato}, {Marzari}, {Boccato},
  {Claudi}, {Cosentino}, {Covino}, {Gratton}, {Maggio}, {Micela}, {Molinari},
  {Pagano}, {Piotto}, {Poretti}, {Smareglia}, {Affer}, {Biazzo}, {Bignamini},
  {Esposito}, {Giacobbe}, {H{\'e}brard}, {Malavolta}, {Maldonado}, {Mancini},
  {Martinez Fiorenzano}, {Masiero}, {Nascimbeni}, {Pedani}, {Rainer}, \&
  {Scandariato}}]{bon17}
{Bonomo}, A.~S., {Desidera}, S., {Benatti}, S., {et~al.} 2017, \aap, 602, A107

\bibitem[{{Boyajian} {et~al.}(2015){Boyajian}, {von Braun}, {Feiden}, {Huber},
  {Basu}, {Demarque}, {Fischer}, {Schaefer}, {Mann}, {White}, {Maestro},
  {Brewer}, {Lamell}, {Spada}, {L{\'o}pez-Morales}, {Ireland}, {Farrington},
  {van Belle}, {Kane}, {Jones}, {ten Brummelaar}, {Ciardi}, {McAlister},
  {Ridgway}, {Goldfinger}, {Turner}, \& {Sturmann}}]{boy15}
{Boyajian}, T., {von Braun}, K., {Feiden}, G.~A., {et~al.} 2015, \mnras, 447,
  846

\bibitem[{{Casasayas-Barris} {et~al.}(2021){Casasayas-Barris}, {Palle},
  {Stangret}, {Bourrier}, {Tabernero}, {Yan}, {Borsa}, {Allart}, {Zapatero
  Osorio}, {Lovis}, {Sousa}, {Chen}, {Oshagh}, {Santos}, {Pepe}, {Rebolo},
  {Molaro}, {Cristiani}, {Adibekyan}, {Alibert}, {Allende Prieto}, {Bouchy},
  {Demangeon}, {Di Marcantonio}, {D'Odorico}, {Ehrenreich}, {Figueira},
  {G{\'e}nova Santos}, {Gonz{\'a}lez Hern{\'a}ndez}, {Lavie}, {Lillo-Box}, {Lo
  Curto}, {Martins}, {Mehner}, {Micela}, {Nunes}, {Poretti}, {Sozzetti},
  {Su{\'a}rez Mascare{\~n}o}, \& {Udry}}]{cas21}
{Casasayas-Barris}, N., {Palle}, E., {Stangret}, M., {et~al.} 2021, \aap, 647,
  A26

\bibitem[{{Casasayas-Barris} {et~al.}(2020){Casasayas-Barris}, {Pall{\'e}},
  {Yan}, {Chen}, {Luque}, {Stangret}, {Nagel}, {Zechmeister}, {Oshagh},
  {Sanz-Forcada}, {Nortmann}, {Alonso-Floriano}, {Amado}, {Caballero},
  {Czesla}, {Khalafinejad}, {L{\'o}pez-Puertas}, {L{\'o}pez-Santiago},
  {Molaverdikhani}, {Montes}, {Quirrenbach}, {Reiners}, {Ribas},
  {S{\'a}nchez-L{\'o}pez}, \& {Zapatero Osorio}}]{cas20}
{Casasayas-Barris}, N., {Pall{\'e}}, E., {Yan}, F., {et~al.} 2020, \aap, 635,
  A206

\bibitem[{{Charbonneau} {et~al.}(2000){Charbonneau}, {Brown}, {Latham}, \&
  {Mayor}}]{cha00}
{Charbonneau}, D., {Brown}, T.~M., {Latham}, D.~W., \& {Mayor}, M. 2000, \apjl,
  529, L45

\bibitem[{{Charbonneau} {et~al.}(2002){Charbonneau}, {Brown}, {Noyes}, \&
  {Gilliland}}]{cha02}
{Charbonneau}, D., {Brown}, T.~M., {Noyes}, R.~W., \& {Gilliland}, R.~L. 2002,
  \apj, 568, 377

\bibitem[{{Cox}(2000)}]{cox00}
{Cox}, A.~N. 2000, Allen's Astrophysical Quantities (Springer-Verlag)

\bibitem[{{Crossfield} {et~al.}(2012){Crossfield}, {Knutson}, {Fortney},
  {Showman}, {Cowan}, \& {Deming}}]{cro12}
{Crossfield}, I. J.~M., {Knutson}, H., {Fortney}, J., {et~al.} 2012, \apj, 752,
  81

\bibitem[{{Cubillos} {et~al.}(2020){Cubillos}, {Fossati}, {Koskinen}, {Young},
  {Salz}, {France}, {Sreejith}, \& {Haswell}}]{cub20}
{Cubillos}, P.~E., {Fossati}, L., {Koskinen}, T., {et~al.} 2020, \aj, 159, 111

\bibitem[{{Deming} {et~al.}(2005){Deming}, {Seager}, {Richardson}, \&
  {Harrington}}]{dem05}
{Deming}, D., {Seager}, S., {Richardson}, L.~J., \& {Harrington}, J. 2005,
  \nat, 434, 740

\bibitem[{{Demory} {et~al.}(2013){Demory}, {de Wit}, {Lewis}, {Fortney},
  {Zsom}, {Seager}, {Knutson}, {Heng}, {Madhusudhan}, {Gillon}, {Barclay},
  {Desert}, {Parmentier}, \& {Cowan}}]{bri13}
{Demory}, B.-O., {de Wit}, J., {Lewis}, N., {et~al.} 2013, \apjl, 776, L25

\bibitem[{{Demory} {et~al.}(2011){Demory}, {Seager}, {Madhusudhan}, {Kjeldsen},
  {Christensen-Dalsgaard}, {Gillon}, {Rowe}, {Welsh}, {Adams}, {Dupree},
  {McCarthy}, {Kulesa}, {Borucki}, \& {Koch}}]{bri11}
{Demory}, B.-O., {Seager}, S., {Madhusudhan}, N., {et~al.} 2011, \apjl, 735,
  L12

\bibitem[{{Esteves} {et~al.}(2015){Esteves}, {De Mooij}, \&
  {Jayawardhana}}]{est15}
{Esteves}, L.~J., {De Mooij}, E. J.~W., \& {Jayawardhana}, R. 2015, \apj, 804,
  150

\bibitem[{{Evans} {et~al.}(2015){Evans}, {Aigrain}, {Gibson}, {Barstow},
  {Amundsen}, {Tremblin}, \& {Mourier}}]{eva15}
{Evans}, T.~M., {Aigrain}, S., {Gibson}, N., {et~al.} 2015, \mnras, 451, 680

\bibitem[{{Evans} {et~al.}(2013){Evans}, {Pont}, {Sing}, {Aigrain}, {Barstow},
  {D{\'e}sert}, {Gibson}, {Heng}, {Knutson}, \& {Lecavelier des
  Etangs}}]{eva13}
{Evans}, T.~M., {Pont}, F., {Sing}, D.~K., {et~al.} 2013, \apjl, 772, L16

\bibitem[{{Gao} {et~al.}(2020){Gao}, {Thorngren}, {Lee}, {Fortney}, {Morley},
  {Wakeford}, {Powell}, {Stevenson}, \& {Zhang}}]{gao20}
{Gao}, P., {Thorngren}, D.~P., {Lee}, E. K.~H., {et~al.} 2020, Nature
  Astronomy, 4, 951

\bibitem[{{Grimm} \& {Heng}(2015)}]{gri15}
{Grimm}, S.~L. \& {Heng}, K. 2015, \apj, 808, 182

\bibitem[{{Grimm} {et~al.}(2021){Grimm}, {Malik}, {Kitzmann},
  {Guzm{\'a}n-Mesa}, {Hoeijmakers}, {Fisher}, {Mendon{\c{c}}a}, {Yurchenko},
  {Tennyson}, {Alesina}, {Buchschacher}, {Burnier}, {Segransan}, {Kurucz}, \&
  {Heng}}]{gri21}
{Grimm}, S.~L., {Malik}, M., {Kitzmann}, D., {et~al.} 2021, \apjs, 253, 30

\bibitem[{{Hapke}(1981)}]{hap81}
{Hapke}, B. 1981, \jgr, 86, 3039

\bibitem[{{Heng} \& {Demory}(2013)}]{hen13}
{Heng}, K. \& {Demory}, B.-O. 2013, \apj, 777, 100

\bibitem[{{Heng} {et~al.}(2021){Heng}, {Morris}, \& {Kitzmann}}]{hen21}
{Heng}, K., {Morris}, B.~M., \& {Kitzmann}, D. 2021, Nature Astronomy, 5, 1001

\bibitem[{{Heng} \& {Tsai}(2016)}]{hen16}
{Heng}, K. \& {Tsai}, S.-M. 2016, \apj, 829, 104

\bibitem[{{Henry} {et~al.}(2000){Henry}, {Marcy}, {Butler}, \& {Vogt}}]{hen00}
{Henry}, G.~W., {Marcy}, G.~W., {Butler}, R.~P., \& {Vogt}, S.~S. 2000, \apjl,
  529, L41

\bibitem[{{Hooton} {et~al.}(2021){Hooton}, {Hoyer}, {Kitzmann}, {Morris},
  {Smith}, {Collier Cameron}, {Futyan}, {Maxted}, {Queloz}, {Demory}, {Heng},
  {Lendl}, {Cabrera}, {Csizmadia}, {Deline}, {Parviainen}, {Salmon}, {Sulis},
  {Wilson}, {Bonfanti}, {Brandeker}, {Demangeon}, {Oshagh}, {Persson},
  {Scandariato}, {Alibert}, {Alonso}, {Anglada Escud{\'e}}, {B{\'a}rczy},
  {Barrado}, {Barros}, {Baumjohann}, {Beck}, {Beck}, {Benz}, {Billot},
  {Bonfils}, {Bourrier}, {Broeg}, {Busch}, {Charnoz}, {Davies}, {Deleuil},
  {Delrez}, {Ehrenreich}, {Erikson}, {Farinato}, {Fortier}, {Fossati},
  {Fridlund}, {Gandolfi}, {Gillon}, {G{\"u}del}, {Isaak}, {Jones}, {Kiss},
  {Laskar}, {Lecavelier des Etangs}, {Lovis}, {Luntzer}, {Magrin},
  {Nascimbeni}, {Olofsson}, {Ottensamer}, {Pagano}, {Pall{\'e}}, {Peter},
  {Piotto}, {Pollacco}, {Ragazzoni}, {Rando}, {Ratti}, {Rauer}, {Ribas},
  {Santos}, {S{\'e}gransan}, {Simon}, {Sousa}, {Steller}, {Szab{\'o}},
  {Thomas}, {Udry}, {Ulmer}, {Van Grootel}, \& {Walton}}]{hoo21}
{Hooton}, M.~J., {Hoyer}, S., {Kitzmann}, D., {et~al.} 2021, arXiv e-prints,
  arXiv:2109.05031

\bibitem[{{Hoyer} {et~al.}(2020){Hoyer}, {Guterman}, {Demangeon}, {Sousa},
  {Deleuil}, {Meunier}, \& {Benz}}]{hoy20}
{Hoyer}, S., {Guterman}, P., {Demangeon}, O., {et~al.} 2020, \aap, 635, A24

\bibitem[{Hunter(2007)}]{hun07}
Hunter, J.~D. 2007, Computing in Science \& Engineering, 9, 90

\bibitem[{{Jensen} {et~al.}(2011){Jensen}, {Redfield}, {Endl}, {Cochran},
  {Koesterke}, \& {Barman}}]{jen11}
{Jensen}, A.~G., {Redfield}, S., {Endl}, M., {et~al.} 2011, \apj, 743, 203

\bibitem[{{Kipping} \& {Bakos}(2011)}]{kip11}
{Kipping}, D. \& {Bakos}, G. 2011, \apj, 730, 50

\bibitem[{{Knutson} {et~al.}(2007){Knutson}, {Charbonneau}, {Noyes}, {Brown},
  \& {Gilliland}}]{knu07}
{Knutson}, H.~A., {Charbonneau}, D., {Noyes}, R.~W., {Brown}, T.~M., \&
  {Gilliland}, R.~L. 2007, \apj, 655, 564

\bibitem[{{Kurucz} \& {Bell}(1995)}]{kur95}
{Kurucz}, R. \& {Bell}, B. 1995, Atomic Line Data (R.L.~Kurucz and B.~Bell)
  Kurucz CD-ROM No.~23.~Cambridge, Mass.: Smithsonian Astrophysical
  Observatory, 1995., 23

\bibitem[{{Lecavelier Des Etangs} {et~al.}(2008){Lecavelier Des Etangs},
  {Vidal-Madjar}, {D{\'e}sert}, \& {Sing}}]{lec08}
{Lecavelier Des Etangs}, A., {Vidal-Madjar}, A., {D{\'e}sert}, J.~M., \&
  {Sing}, D. 2008, \aap, 485, 865

\bibitem[{{Lendl} {et~al.}(2020){Lendl}, {Csizmadia}, {Deline}, {Fossati},
  {Kitzmann}, {Heng}, {Hoyer}, {Salmon}, {Benz}, {Broeg}, {Ehrenreich},
  {Fortier}, {Queloz}, {Bonfanti}, {Brandeker}, {Collier Cameron}, {Delrez},
  {Garcia Mu{\~n}oz}, {Hooton}, {Maxted}, {Morris}, {Van Grootel}, {Wilson},
  {Alibert}, {Alonso}, {Asquier}, {Bandy}, {B{\'a}rczy}, {Barrado}, {Barros},
  {Baumjohann}, {Beck}, {Beck}, {Bekkelien}, {Bergomi}, {Billot}, {Biondi},
  {Bonfils}, {Bourrier}, {Busch}, {Cabrera}, {Cessa}, {Charnoz}, {Chazelas},
  {Corral Van Damme}, {Davies}, {Deleuil}, {Demangeon}, {Demory}, {Erikson},
  {Farinato}, {Fridlund}, {Futyan}, {Gandolfi}, {Gillon}, {Guterman}, {Hasiba},
  {Hernandez}, {Isaak}, {Kiss}, {Kuntzer}, {Lecavelier des Etangs},
  {L{\"u}ftinger}, {Laskar}, {Lovis}, {Magrin}, {Malvasio}, {Marafatto},
  {Michaelis}, {Munari}, {Nascimbeni}, {Olofsson}, {Ottacher}, {Ottensamer},
  {Pagano}, {Pall{\'e}}, {Peter}, {Piazza}, {Piotto}, {Pollacco}, {Ratti},
  {Rauer}, {Ragazzoni}, {Rando}, {Ribas}, {Rieder}, {Rohlfs}, {Safa}, {Santos},
  {Scandariato}, {S{\'e}gransan}, {Simon}, {Singh}, {Smith}, {Sordet}, {Sousa},
  {Steller}, {Szab{\'o}}, {Thomas}, {Tschentscher}, {Udry}, {Viotto}, {Walter},
  {Walton}, {Wildi}, \& {Wolter}}]{len20}
{Lendl}, M., {Csizmadia}, S., {Deline}, A., {et~al.} 2020, \aap, 643, A94

\bibitem[{{Line} {et~al.}(2016){Line}, {Stevenson}, {Bean}, {Desert},
  {Fortney}, {Kreidberg}, {Madhusudhan}, {Showman}, \& {Diamond-Lowe}}]{lin16}
{Line}, M.~R., {Stevenson}, K.~B., {Bean}, J., {et~al.} 2016, \aj, 152, 203

\bibitem[{{Lodders}(2003)}]{lod03}
{Lodders}, K. 2003, \apj, 591, 1220

\bibitem[{{Maxted} {et~al.}(2021){Maxted}, {Ehrenreich}, {Wilson}, {Alibert},
  {Collier Cameron}, {Hoyer}, {Sousa}, {Olofsson}, {Bekkelien}, {Deline},
  {Delrez}, {Bonfanti}, {Borsato}, {Alonso}, {Anglada Escud{\'e}}, {Barrado},
  {Barros}, {Baumjohann}, {Beck}, {Beck}, {Benz}, {Billot}, {Biondi},
  {Bonfils}, {Brandeker}, {Broeg}, {B{\'a}rczy}, {Cabrera}, {Charnoz}, {Corral
  Van Damme}, {Csizmadia}, {Davies}, {Deleuil}, {Demangeon}, {Demory},
  {Erikson}, {Flor{\'e}n}, {Fortier}, {Fossati}, {Fridlund}, {Futyan},
  {Gandolfi}, {Gillon}, {Guedel}, {Guterman}, {Heng}, {Isaak}, {Kiss},
  {Laskar}, {Lecavelier des Etangs}, {Lendl}, {Lovis}, {Magrin}, {Nascimbeni},
  {Ottensamer}, {Pagano}, {Pall{\'e}}, {Peter}, {Piotto}, {Pollacco},
  {Pozuelos}, {Queloz}, {Ragazzoni}, {Rando}, {Rauer}, {Reimers}, {Ribas},
  {Santos}, {Scandariato}, {Simon}, {Smith}, {Steller}, {Swayne}, {Szab{\'o}},
  {S{\'e}gransan}, {Thomas}, {Udry}, {Van Grootel}, \& {Walton}}]{max21}
{Maxted}, P.~F.~L., {Ehrenreich}, D., {Wilson}, T.~G., {et~al.} 2021, \mnras
  [\eprint[arXiv]{2111.08828}]

\bibitem[{{Molli{\`e}re} {et~al.}(2015){Molli{\`e}re}, {van Boekel},
  {Dullemond}, {Henning}, \& {Mordasini}}]{mol15}
{Molli{\`e}re}, P., {van Boekel}, R., {Dullemond}, C., {Henning}, T., \&
  {Mordasini}, C. 2015, \apj, 813, 47

\bibitem[{{Morris} {et~al.}(2021){Morris}, {Heng}, {Brandeker}, {Swan}, \&
  {Lendl}}]{mor21}
{Morris}, B.~M., {Heng}, K., {Brandeker}, A., {Swan}, A., \& {Lendl}, M. 2021,
  \aap, 651, L12

\bibitem[{{Polyansky} {et~al.}(2018){Polyansky}, {Kyuberis}, {Zobov},
  {Tennyson}, {Yurchenko}, \& {Lodi}}]{pol18}
{Polyansky}, O.~L., {Kyuberis}, A.~A., {Zobov}, N.~F., {et~al.} 2018, \mnras,
  480, 2597

\bibitem[{{Rodrigo} \& {Solano}(2020)}]{rod20}
{Rodrigo}, C. \& {Solano}, E. 2020, in XIV.0 Scientific Meeting (virtual) of
  the Spanish Astronomical Society, 182

\bibitem[{{Rodrigo} {et~al.}(2012){Rodrigo}, {Solano}, \& {Bayo}}]{rod12}
{Rodrigo}, C., {Solano}, E., \& {Bayo}, A. 2012, {SVO Filter Profile Service
  Version 1.0}, IVOA Working Draft 15 October 2012

\bibitem[{{Rowe} {et~al.}(2006){Rowe}, {Matthews}, {Seager}, {Kuschnig},
  {Guenther}, {Moffat}, {Rucinski}, {Sasselov}, {Walker}, \& {Weiss}}]{row06}
{Rowe}, J.~F., {Matthews}, J.~M., {Seager}, S., {et~al.} 2006, \apj, 646, 1241

\bibitem[{{Rowe} {et~al.}(2008){Rowe}, {Matthews}, {Seager}, {Miller-Ricci},
  {Sasselov}, {Kuschnig}, {Guenther}, {Moffat}, {Rucinski}, {Walker}, \&
  {Weiss}}]{row08}
{Rowe}, J.~F., {Matthews}, J.~M., {Seager}, S., {et~al.} 2008, \apj, 689, 1345

\bibitem[{{Russell}(1916)}]{rus16}
{Russell}, H.~N. 1916, \apj, 43, 173

\bibitem[{{Santos} {et~al.}(2020){Santos}, {Cristo}, {Demangeon}, {Oshagh},
  {Allart}, {Barros}, {Borsa}, {Bourrier}, {Casasayas-Barris}, {Ehrenreich},
  {Faria}, {Figueira}, {Martins}, {Micela}, {Pall{\'e}}, {Sozzetti},
  {Tabernero}, {Zapatero Osorio}, {Pepe}, {Cristiani}, {Rebolo}, {Adibekyan},
  {Allende Prieto}, {Alibert}, {Bouchy}, {Cabral}, {Dekker}, {Di Marcantonio},
  {D'Odorico}, {Dumusque}, {Gonz{\'a}lez Hern{\'a}ndez}, {Lavie}, {Lo Curto},
  {Lovis}, {Manescau}, {Martins}, {M{\'e}gevand}, {Mehner}, {Molaro}, {Nunes},
  {Poretti}, {Riva}, {Sousa}, {Su{\'a}rez Mascare{\~n}o}, \& {Udry}}]{san20}
{Santos}, N.~C., {Cristo}, E., {Demangeon}, O., {et~al.} 2020, \aap, 644, A51

\bibitem[{{Seager}(2010)}]{sea10}
{Seager}, S. 2010, {Exoplanet Atmospheres: Physical Processes} (Princeton
  University Press)

\bibitem[{{Seager} \& {Sasselov}(2000)}]{sea00}
{Seager}, S. \& {Sasselov}, D.~D. 2000, \apj, 537, 916

\bibitem[{Shapiro \& Wilk(1965)}]{sha65}
Shapiro, S.~S. \& Wilk, M.~B. 1965, Biometrika, 52, 591

\bibitem[{{Sing} {et~al.}(2016){Sing}, {Fortney}, {Nikolov}, {Wakeford},
  {Kataria}, {Evans}, {Aigrain}, {Ballester}, {Burrows}, {Deming},
  {D{\'e}sert}, {Gibson}, {Henry}, {Huitson}, {Knutson}, {Lecavelier Des
  Etangs}, {Pont}, {Showman}, {Vidal-Madjar}, {Williamson}, \&
  {Wilson}}]{sin16}
{Sing}, D.~K., {Fortney}, J.~J., {Nikolov}, N., {et~al.} 2016, \nat, 529, 59

\bibitem[{{Sing} {et~al.}(2008){Sing}, {Vidal-Madjar}, {D{\'e}sert},
  {Lecavelier des Etangs}, \& {Ballester}}]{sin08}
{Sing}, D.~K., {Vidal-Madjar}, A., {D{\'e}sert}, J.~M., {Lecavelier des
  Etangs}, A., \& {Ballester}, G. 2008, \apj, 686, 658

\bibitem[{{Snellen} {et~al.}(2008){Snellen}, {Albrecht}, {de Mooij}, \& {Le
  Poole}}]{sne08}
{Snellen}, I.~A.~G., {Albrecht}, S., {de Mooij}, E.~J.~W., \& {Le Poole}, R.~S.
  2008, \aap, 487, 357

\bibitem[{{Sousa} {et~al.}(2021){Sousa}, {Adibekyan}, {Delgado-Mena}, {Santos},
  {Rojas-Ayala}, {Soares}, {Legoinha}, {Ulmer-Moll}, {Camacho}, {Barros},
  {Demangeon}, {Hoyer}, {Israelian}, {Mortier}, {Tsantaki}, \&
  {Monteiro}}]{sou21}
{Sousa}, S.~G., {Adibekyan}, V., {Delgado-Mena}, E., {et~al.} 2021, \aap, 656,
  A53

\bibitem[{{Southworth}(2010)}]{sou10}
{Southworth}, J. 2010, \mnras, 408, 1689

\bibitem[{{Stassun} {et~al.}(2017){Stassun}, {Collins}, \& {Gaudi}}]{sta17}
{Stassun}, K.~G., {Collins}, K.~A., \& {Gaudi}, B.~S. 2017, \aj, 153, 136

\bibitem[{{Stevenson}(2016)}]{ste16}
{Stevenson}, K.~B. 2016, \apjl, 817, L16

\bibitem[{{Sudarsky} {et~al.}(2000){Sudarsky}, {Burrows}, \& {Pinto}}]{sud00}
{Sudarsky}, D., {Burrows}, A., \& {Pinto}, P. 2000, \apj, 538, 885

\bibitem[{{Swain} {et~al.}(2009){Swain}, {Vasisht}, {Tinetti}, {Bouwman},
  {Chen}, {Yung}, {Deming}, \& {Deroo}}]{swa09}
{Swain}, M.~R., {Vasisht}, G., {Tinetti}, G., {et~al.} 2009, \apjl, 690, L114

\bibitem[{{Szab{\'o}} {et~al.}(2021){Szab{\'o}}, {Gandolfi}, {Brandeker},
  {Csizmadia}, {Garai}, {Billot}, {Broeg}, {Ehrenreich}, {Fortier}, {Fossati},
  {Hoyer}, {Kiss}, {Lecavelier des Etangs}, {Maxted}, {Ribas}, {Alibert},
  {Alonso}, {Anglada Escud{\'e}}, {B{\'a}rczy}, {Barros}, {Barrado},
  {Baumjohann}, {Beck}, {Beck}, {Bekkelien}, {Bonfils}, {Benz}, {Borsato},
  {Busch}, {Cabrera}, {Charnoz}, {Collier Cameron}, {Van Damme}, {Davies},
  {Delrez}, {Deleuil}, {Demangeon}, {Demory}, {Erikson}, {Fridlund}, {Futyan},
  {Garc{\'\i}a Mu{\~n}oz}, {Gillon}, {Guedel}, {Guterman}, {Heng}, {Isaak},
  {Lacedelli}, {Laskar}, {Lendl}, {Lovis}, {Luntzer}, {Magrin}, {Nascimbeni},
  {Olofsson}, {Osborn}, {Ottensamer}, {Pagano}, {Pall{\'e}}, {Peter}, {Piazza},
  {Piotto}, {Pollacco}, {Queloz}, {Ragazzoni}, {Rando}, {Rauer}, {Santos},
  {Scandariato}, {S{\'e}gransan}, {Serrano}, {Sicilia}, {Simon}, {Smith},
  {Sousa}, {Steller}, {Thomas}, {Udry}, {Van Grootel}, {Walton}, \&
  {Wilson}}]{sza21}
{Szab{\'o}}, G.~M., {Gandolfi}, D., {Brandeker}, A., {et~al.} 2021, \aap, 654,
  A159

\bibitem[{{Torres} {et~al.}(2008){Torres}, {Winn}, \& {Holman}}]{tor08}
{Torres}, G., {Winn}, J.~N., \& {Holman}, M.~J. 2008, \apj, 677, 1324

\bibitem[{{Uns{\"o}ld}(1955)}]{uns55}
{Uns{\"o}ld}, A. 1955, {Physik der Sternatmosph{\"a}ren, mit besonderer
  Ber{\"u}cksichtigung der Sonne.} ({Julius Springer, Berlin})

\bibitem[{{Vidal-Madjar} {et~al.}(2004){Vidal-Madjar}, {D{\'e}sert},
  {Lecavelier des Etangs}, {H{\'e}brard}, {Ballester}, {Ehrenreich}, {Ferlet},
  {McConnell}, {Mayor}, \& {Parkinson}}]{vid04}
{Vidal-Madjar}, A., {D{\'e}sert}, J.~M., {Lecavelier des Etangs}, A., {et~al.}
  2004, \apjl, 604, L69

\bibitem[{{Vidal-Madjar} {et~al.}(2003){Vidal-Madjar}, {Lecavelier des Etangs},
  {D{\'e}sert}, {Ballester}, {Ferlet}, {H{\'e}brard}, \& {Mayor}}]{vid03}
{Vidal-Madjar}, A., {Lecavelier des Etangs}, A., {D{\'e}sert}, J.~M., {et~al.}
  2003, \nat, 422, 143

\bibitem[{{Wong} {et~al.}(2021){Wong}, {Kitzmann}, {Shporer}, {Heng},
  {Fetherolf}, {Benneke}, {Daylan}, {Kane}, {Vanderspek}, {Seager}, {Winn},
  {Jenkins}, \& {Ting}}]{won21}
{Wong}, I., {Kitzmann}, D., {Shporer}, A., {et~al.} 2021, \aj, 162, 127

\bibitem[{{Wong} {et~al.}(2020){Wong}, {Shporer}, {Daylan}, {Benneke},
  {Fetherolf}, {Kane}, {Ricker}, {Vanderspek}, {Latham}, {Winn}, {Jenkins},
  {Boyd}, {Glidden}, {Goeke}, {Sha}, {Ting}, \& {Yahalomi}}]{won20}
{Wong}, I., {Shporer}, A., {Daylan}, T., {et~al.} 2020, \aj, 160, 155

\bibitem[{{Zellem} {et~al.}(2014){Zellem}, {Lewis}, {Knutson}, {Griffith},
  {Showman}, {Fortney}, {Cowan}, {Agol}, {Burrows}, {Charbonneau}, {Deming},
  {Laughlin}, \& {Langton}}]{zel14}
{Zellem}, R.~T., {Lewis}, N.~K., {Knutson}, H.~A., {et~al.} 2014, \apj, 790, 53

\end{thebibliography}
\begin{appendix}
\section{Assumed and derived parameters}

\begin{table*}
        \caption{Assumed and derived parameters.}
        \label{tab:params}
        \centering
        \begin{tabular}{llcrl}
                \hline\hline
                Parameters from literature & Symbol & Value & Units & References / Comment\\
                \hline
                Period & $P_0$ & $3.52474859$ & days & 1\\
                Transit time & BJD$_0$ & 2\,452\,826.629283 & days & 2\\
                Planet-to-star radii ratio & $R_p/R_\star$ & ${0.12175}\pm{0.00011}$ & - & 3, 4, 5, 6\\
                Normalised semi-major axis & $a/R_\star$ & ${8.807}\pm{0.051}$ & - & 1, 3, 6, 7 \\
                Orbital inclination & $i$ & ${86.744} \pm {0.022}$ &  deg &  3, 6, 7, 8 \\
                Transit depth & $D$ & $14822 \pm 28$ & ppm & Calculated from $R_p/R_\star$ \\
                Transit duration & $T_{14}$ & $0.12840 \pm 0.00096$ & days & Calc.\ from $P_0$, $a/R_\star$, $i$, and $R_p/R_\star$ \\
                \hline\hline
                Fitted parameter priors & Symbol & Prior$^*$ & Units & Comment\\
                \hline
        Occultation phase width & $W$ & $\mathcal{N}\!\left(0.03643, 0.00054\right)$ & & $W = T_{14}/P_0$  \\
        Occultation depth & $L$ & $\mathcal{U}\!\left(1, 200\right)$ & ppm &  Except visit 5$^\dagger$\\
        Impact parameter & $b$ & $\mathcal{N}\!\left(0.5002, 0.0090\right)$ &  & Calculated from $a/R_\star$ and $i$\\
                Time of inferior conjunction$^\ddagger$ & $T_0$ & $\mathcal{N}\!\left(T_0, 0.01\right)$ & days & Calculated from $P_0$ and BJD$_0$ \\
                \hline\hline
                Fitted parameter posteriors & Symbol & Posterior & Units & Comment\\
                \hline
        Occultation phase width & $W$ & $0.0366 \pm 0.0003 $ &  & \\
        Occultation depth & $L$ & $20.4 \pm 3.3$  & ppm & \\
        Impact parameter & $b$ & $0.5000 \pm 0.007$  &  & \\
                Time of inferior conjunction$^\ddagger$ & $T_0$ & $0.156 \pm 0.003$ & days & Add 2\,459\,453 for BJD\\
                \hline\hline
        \end{tabular}
        \tablefoot{The table shows parameters fixed from the literature and parameters fit using priors, with their posteriors. For the parameters with multiple references, we computed a weighted average. References: (1) \citet{sta17}, (2) \citet{bon17}, (3) \citet{eva15}, (4) \citet{boy15}, (5) \citet{alb12}, (6) \citet{tor08}, (7) \citet{sou10}, (8) \citet{knu07}. 
                \tablefoottext{$*$}{$\mathcal{N}$ and $\mathcal{U}$ denote normal and uniform distributions, respectively.}
                \tablefoottext{$\dagger$}{For visit 5, the occultation depth was derived with an unconstraining prior $\mathcal{U}\!\left(-\infty, \infty\right)$ because the MCMC analysis did not converge with the constrained one.}
                \tablefoottext{$\ddagger$}{Also called mid-transit epoch.}
        }
\end{table*}


\begin{figure*}
\includegraphics[width=14cm]{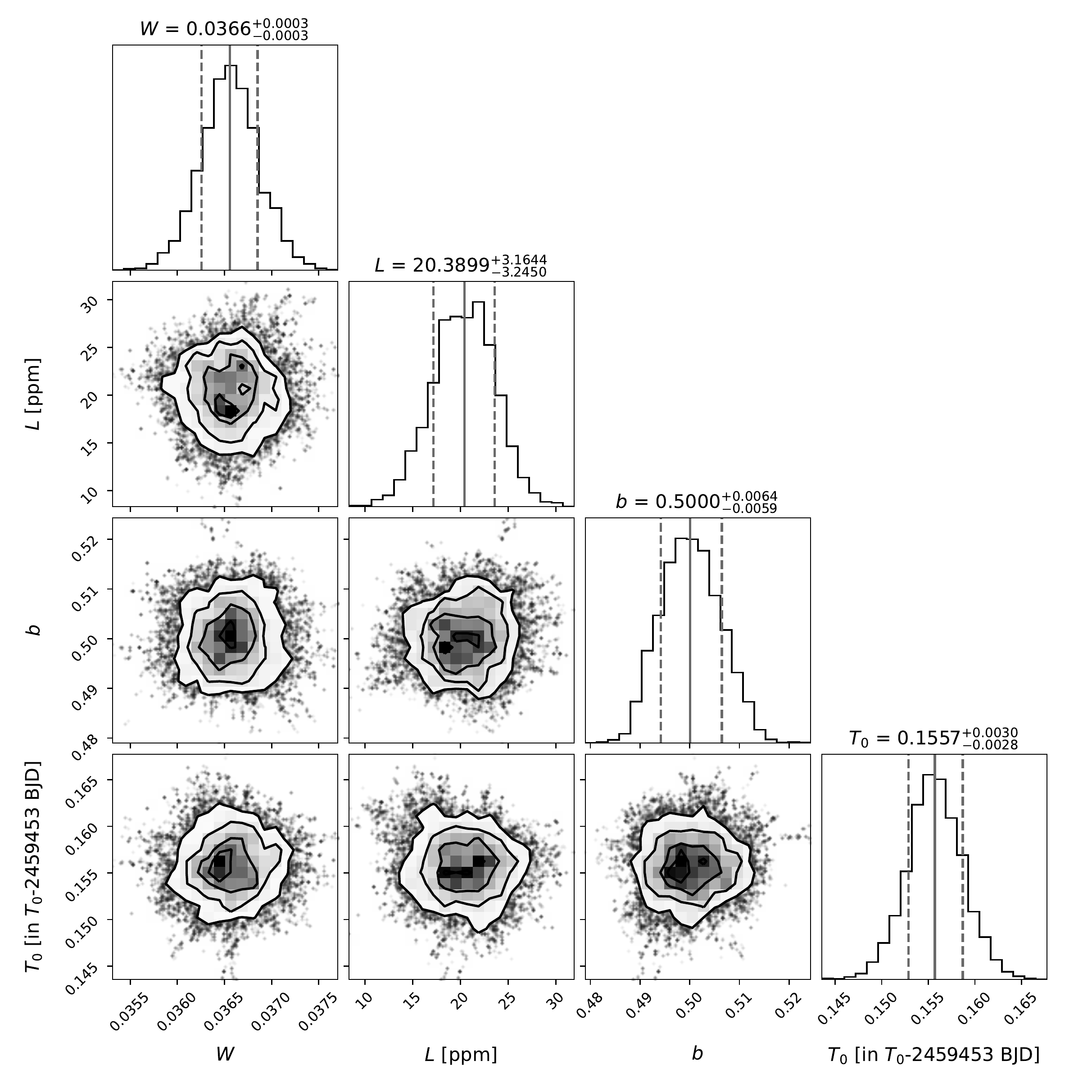}
\vspace{-0.2in}
\caption{Correlation plots for the fitted parameters.}
\label{fig:corner}
\end{figure*}

\end{appendix}

\end{document}